# Studies of local jamming via penetration of a granular medium

M. B. Stone, R. Barry, D. P. Bernstein, M. D. Pelc, Y. K. Tsui, and P. Schiffer[*]

*Department of Physics and Materials Research Institute, Pennsylvania State University, University Park, PA 16802 USA*

We present a series of measurements examining the penetration force required to push a flat plate vertically through a dense granular medium, focusing in particular on the effects of the bottom boundary of the vessel containing the medium. Our data demonstrate that the penetration force near the bottom is strongly affected by the surface properties of the bottom boundary, even many grain diameters above the bottom. Furthermore, the data indicate an intrinsic length scale for the interaction of the penetrating plate with the vessel bottom via the medium. This length scale, which corresponds to the extent of local jamming induced by the penetrating plate, has a square root dependence both upon the plate radius and the ambient granular stress near the bottom boundary, but it is independent of penetration velocity and grain diameter.

[*]schiffer@phys.psu.edu

## I. INTRODUCTION

The propagation of stress through a granular medium presents a complex problem which has been the subject of extensive recent study [1]. An applied stress results in the development of an internal structure resisting the stress, a so-called 'jammed' state [2]. The detailed nature of the jammed state in a granular medium is complicated by the fact that the forces do not propagate uniformly through a granular sample but are localized along directional force chains, and the jammed state is ultimately characterized by the properties of the network of these chains. Jamming can be associated with the stress induced by the grains' own weight, but it can also be induced by the application of a local stress to a dense granular medium. This local jamming has previously been studied in three dimensions through measurements of the drag force on a solid object slowly being pulled horizontally through the medium [3]. The drag force is directly related to jamming since the low velocity drag in a granular medium originates in the force required to break through the jammed grains impeding motion. These measurements demonstrated that, while much of the average drag force could be understood within a simple mean field picture, the stick-slip fluctuations had complex behavior which was apparently affected by the finite size of the vessel containing the medium.

In order to further study boundary effects on the local jamming of granular media, we have performed measurements of the penetration force required to push a flat plate vertically into a granular medium. Specifically, our measurements examine the effects of the bottom boundary of the containing vessel on the penetration force, which reflects how the boundary of the granular sample affects the local jamming induced by penetration. We find that the surface properties of the vessel bottom boundary (e.g. the texture and the friction coefficient) strongly affect the penetration force near the bottom. The results also indicate that there is a characteristic length scale associated with the boundary influence on the penetration force and thus also with the jammed state from which the force originates. This length scale has an unusual square root dependence on the size of the penetrating plate and the ambient granular stress which cannot be easily understood within simple models of elastic or plastic response [4].

## II. EXPERIMENTAL METHODS

Our measurements employed a granular penetrometer shown in Fig. 1(a) consisting of an actuator moving a rod with an attached plate downward through a granular sample contained in a cylindrical vessel. The diameter of the rod (6.4 mm) was smaller than the diameter of the penetrating plates, and was shown to have little impact on the force measurements. A force cell placed between the actuator and the rod measured the resistance force associated with penetration through the medium. Similar instruments are used for soil characterization in engineering applications [5,6] and materials research [7,8,9,10], although these studies are typically designed to study the penetration resistance in the regime where boundary conditions are not important. Such instruments also typically penetrate the media with a sharp cone as opposed to a flat plate; the differences between these two geometries are discussed in the Appendix.

All of the data shown were obtained using the filling procedure illustrated in Fig. 1(b), and each measurement of the penetration resistance was performed on a freshly filled vessel so as to obtain consistent initial conditions. A reservoir of the granular medium being studied was connected to a tube leading from the reservoir to the bottom of the vessel. The outlet diameter of the tube was much smaller than the vessel diameter.

A valve was then opened to allow grains to fill the entire tube leading from the reservoir to the bottom of the vessel. The vessel was then slowly lowered while keeping the tube and the reservoir at a fixed location such that the grains filling the vessel were never under a free-fall condition. The filling was completed by closing the valve and continuing to lower the vessel. The resultant conical surface was not altered except by the penetration. This *cascaded fill* method yielded a consistently lower packing fraction and more consistent force measurements compared to simply filling the container from a point source at a fixed distance above the bottom of the vessel [11,12]. Typical packing fractions for our media were ~59% (for 0.92 mm diameter glass beads). We also performed measurements on samples filled from a fixed point source and measurements where a wire screen or cylindrical sleeve was pulled through the grains after filling. These measurements yielded different magnitudes of the penetration force, and the individual data runs showed larger fluctuations than measurements after using our standard filling method. Data taken with different filling methods were, however, qualitatively equivalent, and the power-law dependence of the length-scales (discussed below) was the same as for our standard filling method.

Measurements of force and position were performed at either 50 Hz or 100 Hz averaging 800 or 400 data points respectively for each individual recorded value at this frequency. These data were then averaged within 1 mm or 0.5 mm distance intervals to reduce fluctuations. The data shown here were further averaged from at least three separate penetration runs. The net force due to the weight of the insertion rod and plate hanging from the load cell was measured prior to entering the medium. This value was then subtracted from all measurements such that the forces shown are due to only the penetration forces.

Our raw data consisted of measurements of the penetration force as a function of height above the bottom of the vessel, $F(z)$, where $z$ is measured from the bottom of the vessel. For each measurement we could vary the diameter of the grains ($d_{grain}$), the diameter of the containing vessel, the velocity of penetration ($v$), the radius of the penetrating plate ($r$), and the amount of filling of the container ($z_{max}$), which is measured from the container bottom to the top of the conical surface created by the filling procedure. Unless otherwise noted, all data shown are for monodisperse spherical glass beads (Jaygo Inc.) with a penetration velocity of $v = 0.5$ mm/sec., container diameter of 213 mm, a support rod diameter of 6.4 mm, and vertical penetrating plate thickness of 7.6 mm. Data were acquired from the top surface ($z \sim z_{max}$) to just above the vessel bottom ($z \sim 4$ mm). We measure the polydispersity of the grains in terms of the standard deviation of the mean diameter measured in a sample of grains, i.e. $d_{grain} = 0.92 \pm 0.04$ mm. The ratio of the mean to the standard deviation in the mean is between 12 and 70 for the range of grain diameters $d_{grain} = 0.38$ mm to $d_{grain} = 3.00$ mm. The playsand medium consists of coarse sand sieved between 20 and 50 mesh (size distribution is approximately 0.3 - 0.85 mm).

## III. DATA AND ANALYSIS

### A. General Characteristics

In Figure 2, we show typical measurements of $F(z)$ for different filling levels of the vessel, $z_{max}$. Each measurement consists of an initial linear regime where the penetration forces appear hydrostatic. For the deepest filling of grains, the hydrostatic regime is followed by a rollover to an almost depth-independent penetration force. This rollover is consistent with the stresses in the granular medium being described by a Janssen-like regime, where the side-walls of the container support a significant portion of the weight of the grains. This feature is also in agreement with recent measurements examining this regime in other quasi-static systems [11]. We label this intrinsic depth dependence of the penetration force (i.e. the force unaffected by the presence of the bottom boundary) as $F_{bulk}(z)$. Near the bottom of the vessel, there is a rapid increase in the penetration force, which is due to the effect of the bottom on the jamming of the grains in front of the penetrating plate. We note that the measurements for the largest values of $z_{max}$, i.e. the deepest filling of grains, do not extend to the bottom of the vessel due to the limited range of motion of the actuator.

Our studies focused on the effect of the bottom of the vessel, i.e. near $z = 0$; however, $F(z)$ in this regime is affected both by the intrinsic depth dependence of the penetration force associated with the stress on the medium due to its weight, and by the proximity of the bottom. To quantify the effect of the proximity of the bottom, two separate measurements were performed, as shown in Fig. 3 where we plot $F(z)$ for $z_{max} = 112$ mm and $z_{max} = 230$ mm. The data for $z_{max} = 230$ mm provide a measure of the penetration force far from the vessel bottom. Thus we can use them to obtain the bulk penetration force, $F_{bulk}(z)$, by translating them to account for different (smaller) values of $z_{max}$. This measurement of $F_{bulk}(z)$ then serves as a measurement of the *background* penetration force, i.e. the depth dependence of the penetration force without the effects of the bottom for a given value of $z_{max}$ (choosing $z_{max}$ considerably smaller than 230 mm).

We use $F_{bulk}(z)$ to directly obtain a measure of the influence of the bottom of the container by examining the differential penetration force $\Delta F(z) = F(z) - F_{bulk}(z)$ as shown in Fig. 3(b).

We plot the differential penetration force, $\Delta F(z)$, in Fig. 3(b) for a vessel with a smooth bottom. A striking feature in these data is the local minimum near the bottom of the vessel. The *negative* value of $\Delta F(z)$ in this regime implies that the presence of the bottom of the container reduces the magnitude of the penetration force relative to moving though the bulk of the medium at the same depth. We probe this effect by altering the surface characteristics of the vessel bottom, specifically by measuring $\Delta F(z)$ for bottoms with different textures (beaded or coated with Teflon) or for bottoms with grooves etched into the surface. Differences in the penetration forces associated with these different surfaces are shown in Fig. 4, and the data illustrate the mechanism producing the local minimum in $\Delta F(z)$. Data from surfaces that impede the radial flow of grains (beaded or concentric circular grooves) have no local minima, while those which do not impede the radial outflow (smooth bottom surface or radial grooves) do have local minima in $\Delta F(z)$. This implies that the local minima are due to the grains sliding more easily along the bottom surface than within the bulk of the material. The sliding mechanism is clearly evident when comparing the measurements for grooved boundary layers as shown in Fig. 4. A bottom boundary textured with concentric circular grooves prevents the radial outflow of grains and has no local minimum in $\Delta F(z)$, but a radial grooved boundary with identical groove widths and depths still allows radial motion of the grains and produces a minimum in $\Delta F(z)$. The minimum in $\Delta F(z)$ is enhanced when the surface friction of the smooth bottom is reduced, as shown in Fig. 4 for a Teflon coated bottom surface. We have varied the size of the concentric grooves from 0.8 mm to 3.2 mm wide, but have been unable to observe any significant systematic change in *F(z)* or *ΔF(z)* for the 0.9 mm diameter beads employed.  The absence of a local minimum in measurements employing the non-spherical playsand medium can be attributed to the rough surfaces and angular nature of those grains not being as conducive to sliding or rolling along the bottom of the vessel.

While the existence or non-existence of a minimum in *ΔF(z)* depends on the surface characteristics of the bottom, for any type of surface there is a sharp rise in *ΔF(z)* as the plate approaches the bottom, i.e. as $z \to 0$.  The nature of the sharp rise is exponential, as demonstrated in Fig. 3 (c), indicating the existence of a length scale associated with the approach of the penetrometer toward the bottom of the vessel.  This exponential can be parameterized by

$$\Delta F(z) = \Phi \exp[-z/\lambda],$$

where $\Phi$ is the amplitude of the differential penetration force and $\lambda$ is a length scale associated with the effect of the bottom of the vessel on the penetration force.  As demonstrated by the straight line in Fig. 3(c), this form typically fits the data over an order of magnitude change of the differential penetration force and to a height of many grain diameters above the bottom of the vessel.  This form fits the data close to the vessel bottom for all grain sizes tested as well as non-spherical media and all the different bottom surfaces of the vessel discussed above, with only a slight increase above exponential behavior observed upon the closest approach to the bottom, i.e. within ~3-5 mm of the bottom.  This rise is presumably due to the trapping of a rigid structure of a few individual grains between the plate and the bottom, rather than any collective behavior.

The length scale $\lambda$ for the effect of the bottom can also be parameterized by the location of the minimum in $\Delta F(z)$, which we label as $\zeta$. The functional form of $\Delta F(z)$ could be fit well by the phenomenological function

$$\Delta F(z) = A\left(1 - \exp\left[\tau(\zeta - z)\right]\right)^2 - A$$

which was used to determine $\zeta$ in a consistent manner. As demonstrated below, $\zeta$ and $\lambda$ are proportional to one another, indicating that they are describing the same physical quantity, i.e. the size of the jammed region which is reorganized as the plate pushes through. In the sections below, we study how this length scale is affected by various physical parameters of the system.

## B. Dependence on System Parameters

### 1. Vessel Diameter Dependence

To examine the influence of the vessel diameter on $F(z)$, we performed measurements with a series of different size vessels as shown in Fig. 5. The penetration force, shown in Fig. 5 (a) for $r = 12.7$ mm, is independent of vessel diameter for diameters greater than approximately 130 mm. This is true for both shallow ($z_{max} = 112$ mm) and deep ($z_{max} = 230$ mm) filling of the vessels. For smaller diameter vessels, the system becomes analogous to a piston where the volume of grains disturbed by the penetrating plate is constrained by the side-walls of the vessel. The differential penetration force, Fig. 5 (b), also converges for vessel diameters greater than approximately 130 mm. The values of $\lambda$ and $\zeta$ are shown in Fig. 5 (c) as a function of vessel diameter. As mentioned above, the values of $\zeta$ and $\lambda$ are proportional to one another (this proportionality is also demonstrated below for variations of other system parameters), and, as expected, they become independent of vessel diameter in the limit of large vessels. Since we want to probe the influence of the bottom on the penetration force without direct interactions with the walls, we employed the relatively large 213 mm diameter vessel for all of the data described below.

**2. Velocity Dependence**

Because we wanted to probe the effects of local jamming of the grains, it was important to study the penetration force in the low velocity limit where the grains are not fluidized in front of the penetrating plate. We expected that the penetration force in this limit should be independent of velocity since it is determined by the amount of force needed to break through a jammed state. The data in Fig. 6 indicate that the penetration force and differential penetration force are indeed independent of velocity over two orders of magnitude of velocity. This result agrees with the velocity independence found in granular drag measurements [3], and is in sharp contrast to fluid drag which originates in momentum transfer to the fluid. We also found that the length scales $\lambda$ and $\zeta$ are independent of velocity as shown in shown in Fig. 6(b) and (c).

Measurements at the lowest velocities indicated the force to be stick-slip in nature, again in analogy to granular drag [3]. Although we do not fully describe this behavior here, we note that this is consistent with the penetration process consisting of a linear increase in force as the penetrometer stresses a jammed granular state followed by a breaking or collapse of this state indicated by a rapid decrease in the penetration force.

### 3. Plate Thickness Dependence

Since the process of penetration requires that the grains below the plate be reorganized and eventually displaced to be above the plate, another parameter we varied was the vertical thickness of the plates. As shown in Fig. 7(a) and (b), the thickness of the penetrating plate has little effect on the nature of the penetration or differential penetration forces. The effect of different plate thicknesses was observed at the largest depths where the stress on the grains (and presumably the frictional force on the sides) is the largest. As demonstrated in Fig. 7(b) and (c), the data near the vessel bottom for smaller $z_{max}$ appear to be unaffected by plate thickness, and the parameters $\lambda$ and $\zeta$ also show no systematic dependence. As noted above, all data presented in other figures were taken with a plate thickness of 7.6 mm.

### 4. Grain Diameter Dependence

As in prior drag force measurements [3], we found that $F(z)$ does not have a systematic dependence upon the grain diameter for our spherical glass beads. This is demonstrated in Fig. 8 for a series of different size monodisperse spherical glass beads (the grain diameter was always much smaller than that of the penetrating plates). While there are differences between the penetration force curves, these can be explained as differences in surface textures and particle size distributions among the different samples. Surprisingly, there is very little dependence of $\lambda$ and $\zeta$ on the diameter of the spherical grains, even though $\lambda$ is not much larger in magnitude than the larger grain sizes. Penetration of the non-spherical playsand did require forces much larger than those for spherical glass media. This is presumably due to a stronger jammed state resulting from the interlocking of the non-spherical grains. Likewise, there is an absence of a local minimum in the differential penetration force and an increase in the value of the length scale $\lambda$ as measured for non-spherical media. This is also likely due to differences in the strength or the extent of the jammed state as compared to spherical media.

## 5. Fill Height Dependence

Another parameter which determines the penetration force near the vessel bottom is the height to which the vessel is filled, $z_{max}$. Increasing $z_{max}$ increases the ambient stress near the vessel bottom which in turn affects the jamming of the grains and thus the penetration force (as can be seen from the data in Fig. 2). The exponential rise of $\Delta F(z)$ can be seen for all values of $z_{max}$ as shown in Fig. 9 where we plot data for two different grain sizes as well as for smooth and circularly grooved vessel bottoms. Note that the curves for the larger values of $z_{max}$ are almost identical, suggesting that these data are in the Janssen regime. In this regime, much of the weight of the grains was supported by the vessel walls, and thus $z_{max}$ was not directly proportional to the weight-induced stress at the vessel bottom. While we could not directly measure the weight-induced stress near the vessel bottom, we expect it to be proportional to the measured quantity $F_0 = F_{bulk}(z = 0)$. We can thus use $F_0$ to examine how $\Delta F$ depends on the ambient stress state of the grains near the bottom of the vessel (note that $F_0$ is obtained from $F_{bulk}$ as indicated in Fig. 3(a) and is not affected by the proximity of the vessel bottom). The values of the length scale $\lambda$ obtained for different measurements are plotted as a function of $F_0$ in Fig. 10. As indicated by the lines on the figure, the length scale varies approximately as $\lambda \propto \sqrt{F_0}$ for both circularly grooved and smooth bottom vessels and for different grain diameters, although the magnitude of $\lambda$ varies somewhat among these different conditions.

## 6. Plate Size Dependence

The final system parameter which we varied in our studies of the penetration force was the radius of the circular penetrating plate, $r$. Typical results with different plate radii are shown in Fig. 11, which clearly demonstrates that the penetration force increases with the size of the penetrating plate. Except near the top surface and bottom boundary of the vessel, $F(z)$ was proportional to the cross sectional area of the penetrating plate. This is shown in Fig. 12, where we plot $F_0$ for $z_{max} = 112$ mm as a function of the penetrating plate area for different grain diameters including the playsand media. This result is consistent with mean field expectations and the data from previous drag force experiments [3, 13]. Square penetrating plates were also used for $d_{grain} = 0.92 \pm 0.04$ mm, and the values of $F_0$ are indistinguishable with those of circular cross section plates as shown in Fig. 12.

We now consider the effect of plate area on the length scales of the jammed state, $\lambda$ and $\zeta$. We plot $\Delta F(z)$ for a range of plate radii for two different grain diameters and both smooth and circularly grooved vessel bottoms in Fig. 13. These data indicate that in varying the diameter of the penetrating plate, the length scale $\lambda$ continually increases with increasing area. The plate area dependence of $\lambda$ can be well fit to a power law, $\lambda \propto A^{1/4}$, for all grain diameters and for circular as well as square plates as illustrated in Fig. 14. The same power law behavior is observed for both circularly grooved and smooth bottoms, although $\lambda$ is somewhat larger for the grooved bottoms than for smooth bottoms as shown in Fig. 8 (c). This is expected, since by reducing the transverse mobility of the grains along the bottom of the vessel the grooved bottom surfaces increases the strength of the jammed state. Fig. 14 also indicates the same $A^{1/4}$ dependence is displayed by $\zeta$, further suggesting that this dependence is a physically robust result.

In order to further understand the dependence of the length scale $\lambda$ on the penetrating plate cross-sectional area, we have modeled the granular assembly as an elastic medium with a simple Coulomb yield criterion in the vicinity of the advancing plate [14]. The granular medium was treated as a semi-infinite elastic medium with a force applied to it over a finite area as shown in Fig. 15(g). This is known as the Boussinesq [15] problem in the field of elasticity and the stresses which exist in such a medium have been analytically solved in three-dimensions for both circular and square cross sections of the applied force [16, 17]. We have calculated the stress as a function of area for a constant value of force per unit area applied to the surface of an elastic medium. The radial and longitudinal stress components, $\sigma_{rr}$ and $\sigma_{zz}$, are plotted in Fig. 15 (a) and (b) respectively. A ratio of unity was used for the values of the Lame coefficients, which are simple functions of the bulk and Young moduli, although the qualitative structure of the stress as a function of location in the elastic medium is similar for other ratios. In analogy to simple static friction, we considered the failure criterion to be when the stress in the radial direction exceeds some fraction of the stress in the vertical direction $\sigma_{rr} \geq \mu \sigma_{zz}$, where $\mu$ represents a friction coefficient. As shown in Fig. 15 (c), this yields a truncated hemispherical volume beneath the applied force for which the slip criterion holds. As the area over which the force is applied is increased with fixed pressure, the locus describing the hemisphere will extend further into the medium. Choosing the point for $\sigma_{rr} \geq \mu \sigma_{zz}$ beneath the center of the applied force as the *slip point,* the purely elastic model describes the distance to the *slip point* as increasing proportionally with the width (or diameter) of the plate, in contrast to our experimental result of $\lambda \propto A^{1/4} \equiv r^{1/2}$. This implies that our penetration measurements describe non-elastic behavior in the vicinity of the vessel bottom.

Photoelastic [18] studies and finite element numerical simulations [19] of penetration in the bulk of the medium both describe a volume of material moving close to the penetrometer in a plastic zone surrounded by larger regions of non-linear and linear elastic behavior [20]. We are also able to analyze the plate-size dependence in analogy with plastic failure criteria associated with bearing capacity often described in the discipline of soil mechanics [24]. We again begin with the geometry shown in Fig. 15(g), where the support footing geometry is identical to our penetration system. The failure region under a support footing has a defined shape and vertical extent which is a function of the size of the footing (i.e. the radius of the penetrating plate) and the angle of repose of the medium [24]. The plastic failure region for an angle of repose of 30 degrees is illustrated in Fig. 15(e). The different bounded regions of the figure are described by three different directions of plastic flow of the medium; we define the slip points C, F and E as the vertical extent of these three regions. We then calculate numerically, based upon formulae describing the dimensions of the plastic regions in terms of the angle of repose [24], the depth of these slip points as a function of the width of the applied force as shown in Fig. 15(f). These results indicate that, as for the purely elastic model, the plastic model describes the depth of the slip point to be directly proportional to the width of the area of the applied force. This again does not correspond in any simple way with our finding that the length scale for the penetration-induced jamming $\lambda \propto A^{1/4} \equiv r^{1/2}$.

## C. Scaling

We have seen that $\lambda$ depends both upon the stress at the bottom of the vessel and the radius of the plate through the relationships $\lambda \propto \sqrt{F_0}$ and $\lambda \propto \sqrt{r}$ when each is varied independently of the other. Since there is an intrinsic radius dependence of $F_0$, $F_0 \propto r^2$ as demonstrated in Fig. 12, the finding of $\lambda \propto \sqrt{r}$ for constant $F_0$ requires that the combined dependence on these two parameters is given by $\lambda \propto \sqrt{F_0/r}$. This scaling is directly demonstrated in Fig. 16 for two grain sizes and both circularly grooved and smooth bottomed vessels. Scaling with this relation causes all of the data in Fig. 9 and Fig. 13 to collapse onto characteristic exponential curves for the particular grain size and vessel bottom as shown in Fig. 17 for a wide range of both $F_0$ and $r$. The data for the $r = 4.4$ mm and $r = 6.5$ mm diameter plates do not collapse, indicating that the scaling only holds when the grains are much smaller than the penetrating plate.

The scaling in Fig. 17 can be summarized by expressing the exponential dependence as $\Delta F(z) = C_1 F_0 e^{-z/\left[C_2 \sqrt{F_0/r}\right]}$, where $\Phi = C_1 F_0$ and $\lambda = C_2 \sqrt{F_0/r}$, where $C_1$ and $C_2$ are constants that incorporate bead roughness, packing, shape, bulk density, gravity, etc. The dependence of $\Phi$ on plate size and $F_0$ might have been predicted simply by dimensional analysis. By contrast, the square root dependence of $\lambda$ is quite unexpected and presents a theoretical challenge. From a dimensional standpoint, including other relevant parameters (i.e. gravity, $g$, and the bulk density of the grain material, $\rho$) suggests that a likely relationship for the length-scale is $\lambda \propto \sqrt{\dfrac{F_0}{rg\rho}}$. Testing of this expression would, however, be difficult in that it would require centrifugal or microgravity environments and significant variations in the bulk density of the medium.

Changing this last property undoubtedly would also change the surface friction between grains in some non-systematic way, making comparisons difficult.

## IV. CONCLUSIONS

Our measurements of the penetration force have two main results. The first is that the penetration force near a solid boundary can be affected by the surface characteristics of the boundary, even when the penetrating object is many grain diameters from the boundary. While it is common knowledge that boundary surface properties affect shear flow of granular media, these measurements directly demonstrate that the surface can also impact the jamming induced by applied stress normal to the boundary. On a qualitative level, the results show that penetration provides a means of "remote" sensing of the surface properties since the penetration force reflects those properties at a considerable distance away from the boundary.

The second main result from our study is the finding of an intrinsic length scale describing the interaction of the penetrating object and the boundary. This result strongly suggests that there is an intrinsic length scale to the locally jammed state created by the local application of pressure by the penetrating object. The size of the jammed state is determined by the particular failure mode of the jammed state, which in turn depends on the structure of the force chains originating from the plate [7,9,10]. Because the force propagation is long-range in nature, the existence of a length scale is somewhat surprising. This observed length scale in the penetration geometry is qualitatively different from the observed length scales for granular shearing in Couette geometries [21,22] which are associated with the decay of transverse velocity rather than the transmission of force normal to the direction of motion.

Our results indicate that penetration studies can provide an excellent probe of the structure of local jamming and the dynamics of the failure of a jammed state. Although our results do not appear to be consistent with simple elastic and plastic failure models, the observed dependence of $\lambda$ on the stress in the medium and the plate size may provide input to simulations to test different force propagation models. While our apparatus is not sensitive to the detailed morphology of the jammed region, this could be probed either through modeling (as has been performed for fluids [23] and conical penetrometers in the large sample limit [5,6]) or through imaging with X-ray tomography or MRI [21]. Studies of the effects of cohesion, lubrication, grain morphology, and polydisperse grain sizes would shed further light on how these technologically important factors affect local jamming. Furthermore, such studies would provide an opportunity to connect the recent advances in the fundamental character of jamming with the accumulated practical understanding of the processes involved with penetration testing.

## APPENDIX: Flat plate vs. conical penetrometers

Typical penetration testing for engineering and soil science applications utilizes the "Standard Conical Penetration Test". These tests use an apparatus similar to ours, but portable so that it is suitable for performing measurements in different locations, and, rather than a penetrating plate, these instruments use a conical tip. We examined the differences in measurements between a cylindrical plate and a cone with its diameter equal in length to its height. The cone and the plate both had a circular cross section with a radius of 12.7 mm. The apex of the cone was used for determining the value of the independent variable $z$, while the bottom surface was used for the penetrating plate. Both measurements have very similar penetration force curves, but the forces for the cone were consistently less than those of the circular plate. This is especially true near the bottom of the vessel. The differential penetration force displays a minimum for cones approaching smooth vessel bottoms, as was the case for the penetrating plates. The apex of the cone does not, however, allow a close approach to the bottom of the vessel, limiting the ability to study boundary effects with this geometry. While a vessel bottom which was sloped to match the angle of the cone would circumvent this difficulty, such a geometry would be sufficiently more complicated that we chose to focus on the flat plate geometry instead.

## ACKNOWLEDGEMENTS

We thank J. R. Banavar, A.-L. Barabási and J. P. Schiffer for helpful discussions. Research was supported by the NASA Microgravity Fluid Physics Program through Grant NAG3-2384.

**Figure Captions**

**Fig. 1.** (Color online) Granular penetrometer apparatus used for penetration measurements. (a) Actuator consists of a high torque (0.44 Nm) computer controlled stepper motor with a 200 mm range of linear motion and a distance dependent voltage divider serving as a measure of the depth of penetration. The force cell has a capacity of 110 N acting in either tension or compression. Measurements are controlled and monitored via computer using a commercial data acquisition board. The entire actuator and penetration assembly is rigidly attached to a wall. $z = 0$ represents the bottom of the vessel containing the grains. (b) Preparation method of the granular sample as described in the text. The descending support is lowered via a mechanical lift such that grains were never in free fall motion. This filling procedure was performed prior to every measurement.

**Fig. 2.** (Color online) Height dependence of penetration force, $F(z)$, for $d_{grain} = 0.92 \pm 0.04$ mm and a $r = 12.7$ mm penetrating plate. The fill height of the vessel, $z_{max}$, increases from left to right for the data shown. The three different indicated regimes of the penetration curves (linear/hydrostatic, wall supported, and bottom dominated) are discussed in the text.

**Fig. 3.** (Color online) Penetration force, $F(z)$, and differential penetration force, $\Delta F(z)$, as a function of height above the bottom of the vessel, $z$, for two different fill heights using $d_{grain} = 0.92 \pm 0.04$ mm diameter glass beads and a 9.5 mm radius penetrating plate. (a) $F(z)$ for $z_{max} = 112$ mm (red open squares) and $F(z)$ for $z_{max} = 230$ mm (green open

triangles). $F_{bulk}(z)$ for $z_{max} = 112$ mm (solid line) is obtained by translating the $F(z)$ data for $z_{max} = 230$ mm to be aligned with $F(z)$ for $z_{max} = 112$ mm near the top grain surface. The location of $F_0 = F_{bulk}(z = 0)$, as described in the text is also indicated. Data are plotted for every third data point for the $z_{max} = 230$ mm measurement. (b) and (c) $\Delta F(z)$ on linear and semilogarithmic scales respectively. Solid lines are fits to Equations 1 and 2 respectively as described in the text. Note that the minimum in $\Delta F(z)$ results from the smooth vessel bottom used for these data.

**Fig. 4.** (Color online) $\Delta F(z)$ for a series of different vessel bottom boundaries ($d_{grain} = 0.92 \pm 0.04$ mm and $r = 12.7$ mm). Vessel bottoms and penetrating plates were machined from aluminum (the beaded bottom was coated with 3 mm diameter glass beads, and the Teflon bottom was covered with a thin Teflon sheet). Circularly grooved boundaries are composed of concentric grooves with a depth and width as indicated. The radial grooved bottom consisted of 16 radial grooves, 0.79 mm wide and 0.79 mm deep. Error bars have been suppressed for presentation and markers are shown for every tenth data-point.

**Fig. 5.** (Color online) Vessel diameter dependence of penetration forces using $d_{grain} = 0.92 \pm 0.04$ mm, $r = 12.7$ mm and a smooth-bottomed vessel. (a) $F(z)$ for a series of different vessel diameters for $z_{max} = 112$ mm (closed symbols) and $z_{max} = 230$ mm (open symbols) fillings of the medium. Points are shown for every twentieth data point and error bars have been suppressed for clarity of figure. (b) $\Delta F$ vs. $z$ plotted using a semilogarithmic scale with color coding identical to panel (a). Data for 54 mm diameter vessel are not shown due to lack of significant exponential behavior. Error bars and points at $z > \zeta$ have been suppressed for clarity of figure. Color coding is identical to panel (a). (c) Parameters $\lambda$ and $\zeta$ as a function of vessel diameter. The right vertical axis for $\zeta$ is plotted using a scale six times larger than the left vertical scale for $\lambda$.

**Fig. 6.** (Color online) Velocity dependence of $F(z)$ and $\Delta F(z)$ for $d_{grain}$ = 0.92 ± 0.04 mm, $r$ = 12.7 mm and a smooth-bottomed vessel. (a) $F(z)$ for a series of velocities for both $z_{max}$ = 112 mm (closed symbols) and $z_{max}$ = 230 mm (open symbols) measurements. Points are shown for every twentieth data point, and error bars have been suppressed for clarity of figure. (b) $\Delta F(z)$ with points shown for every fifth data point, and error bars suppressed for clarity. Color coding is identical to panel (a). (c) Parameters $\lambda$ and $\zeta$ as a function of the penetrating plate velocity plotted on a semilogarithmic scale. The right vertical axis for $\zeta$ is plotted using a scale six times larger than the left vertical scale for $\lambda$.

**Fig. 7.** (Color online) Thickness dependence of the penetrating plate for $F(z)$ and $\Delta F(z)$ as measured for $d_{grain}$ = 0.92 ± 0.04 mm, $r$ = 12.7 mm and a smooth-bottomed vessel. (a) $F(z)$ for both $z_{max}$ = 112 mm (closed symbols) and $z_{max}$ = 230 mm (open symbols) measurements. (b) $\Delta F(z)$ with color coding identical to panel (a). (c) Fitting parameters $\lambda$ (left axis) and $\zeta$ (right axis) as a function of the penetrating plate thickness.

**Fig. 8.** (Color online) Grain diameter dependence of the penetration forces employing spherical glass beads (Jaygo Inc.), $r$ = 12.7 mm and a smooth-bottomed vessel. (a) $F(z)$ for a series of different diameter glass beads and "playsand" media for shallow and deep fillings. Points are shown for every twentieth data point, and error bars have been suppressed for clarity of figure. (b) and (c) $\lambda$ as a function of grain diameter for smooth (open symbols) and circularly grooved (closed symbols) bottom container measurements respectively for a series of different penetrating plate radii. For smooth-bottomed

vessels, the "playsand" data had no local minima and a value of $\lambda = 7.15 \pm 7$ mm for the data shown.

**Fig. 9.** (Color online) $\Delta F(z)$ for a series of different filling depths, $z_{max}$, for $r = 12.7$ mm and $d_{grain} = 0.92 \pm 0.04$ mm (left column) and $d_{grain} = 0.56 \pm 0.04$ mm (right column). For each panel, $z_{max}$ increase from left to right as $z_{max} = 62$ mm, 82 mm, 101 mm, 116 mm, 135 mm, 153 mm, 174 mm and 193 mm. Panels (b) and (d) are results obtained using a circularly grooved bottom with a groove size of 1.6 mm. Note the persistence of the length scale behavior to larger values of $z$ in the grooved bottom data compared to the smooth bottom data.

**Fig. 10.** (Color online) Length scale, $\lambda$, as a function of the stress at the bottom of the vessel, as measured by $F_0$, for $d_{grain} = 0.56 \pm 0.04$ mm (red triangles) and $d_{grain} = 0.92 \pm 0.04$ mm (black squares) using both grooved (closed symbols) and smooth (open symbols) bottom vessels. Solid lines are individual fits to the power-law $\lambda \propto (F_0)^{1/2}$.

**Fig. 11.** (Color online) Penetrating plate radius, $r$, dependence of the penetration force, $F(z)$, using $d_{grain} = 0.92 \pm 0.04$ mm glass beads with $z_{max} = 112$ mm and plotted on a semilogarithmic scale. Points are shown for every tenth data point, and error bars have been suppressed for clarity.

**Fig. 12.** (Color online) $F_0$ measured for $z_{max} = 112$ mm vs. area of penetrating plate for different grain diameters. Solid line is proportional to area of penetrating plate. Square cross-section penetrating plates are included in the $d_{grain} = 0.92 \pm 0.04$ mm measurements. Deviations for the force being proportional to the area for the smallest

penetrating plates ($r = 4.4$ mm) used is likely due to the penetration force being comparable to the drag force of the insertion rod ($r = 3.2$ mm).

**Fig. 13.** (Color online) $\Delta F(z)$ for a series of different radii penetrating plates using both $d_{grain} = 0.92 \pm 0.04$ mm (left column) and $d_{grain} = 0.56 \pm 0.04$ mm (right column). For each panel, the plate radii increase from left to right as $r =$ 6.4 mm, 9.5 mm, 12.7 mm, 15.9 mm, 19.1 mm and 25.4 mm. All data are plotted on a semilogarithmic scale to demonstrate the exponential behavior of $\Delta F(z)$. Panels (b) and (d) are results obtained using a circularly grooved bottom with a groove size of 1.6 mm. Note the persistence of the length scale behavior to larger values of $z$ in the grooved bottomed data. For comparison, error bars are only shown for the $r = 25.4$ mm penetrating plate.

**Fig. 14.** (Color online) Length scales, $\lambda$ (lower points) and $\zeta$ (upper points), as a function of penetrating plate area for different grain diameters and a smooth bottomed vessel. Data are included for both circular and square cross section penetrating plates. Solid lines are individual comparisons to the power-law $\lambda \propto A^{1/4}$ and $\zeta \propto A^{1/4}$.

**Fig. 15.** Elasticity, (a)-(d), and plasticity, (e)-(f), calculations for the application of a force applied at the top of the respective type of medium as shown in the geometry of panel (g). Z is the direction of application of the force over the XY plane. (a) $\sigma_{rr}$. (b) $\sigma_{zz}$. (c) $\sigma_{rr}$ where $\sigma_{rr} > \mu\sigma_{zz}$, using a value of $\mu=0.35$. (d) The distance to the slip point, $z_{slip}$, as a function of the width of the applied force for an elastic medium. Calculations were made for three different ratios of Lame coefficients. Solid lines are linear fits. (e) Calculated failure regions in the x-z plane from plasticity theory for a continuous footing with a finite width in the geometry shown in (g). Plasticity regions are based upon

derivations and figures shown in references [24], and calculated using an angle of repose of 30 degrees. The vertical scale of the figure is twice the horizontal scale. AB is the width of the applied force. Triangle ABC (BDE) is a zone of vertical (horizontal) motion of the medium. The region bounded by BCE is in radial shear. The curve CE is a logarithmic spiral tangent to both AC and ED. The vertical distance to points C, E and F are described as heights h1, h2 and h3 respectively. (f) Heights h1, h2 and h3 as a function of the width of the applied force in the plastic model.

**Fig. 16.** (Color online) Length scale, $\lambda$, as a function of the quantity $\sqrt{F_0/r}$ for circularly grooved (closed symbols) and smooth (open symbols) bottomed vessels for (a) $0.92 \pm 0.04$ mm diameter grains and (b) $0.56 \pm 0.04$ mm diameter grains. Data points include those derived from $z_{max}$ dependence and plate size dependence measurements. Solid lines are linear fits with an intercept at the origin as described in the text.

**Fig. 17.** (Color online) Scaled differential penetration force ($\Delta F \to \Delta F/F_0$) vs. scaled height ($z \to z/(F_0/r)^{1/2}$). Data in each panel are shown for both height and disc radius dependence. (a) and (b) $d_{grain} = 0.92 \pm 0.04$ mm using smooth and circularly grooved bottom vessels respectively. (c) and (d) $d_{grain} = 0.56 \pm 0.04$ using smooth and circularly grooved bottom vessels respectively. Color coding is identical to the data shown in Fig. 13 and Fig. 9. Dotted lines are used for data where the plate size was varied and solid lines are used for data where $z_{max}$ was varied  Error bars are suppressed for clarity.

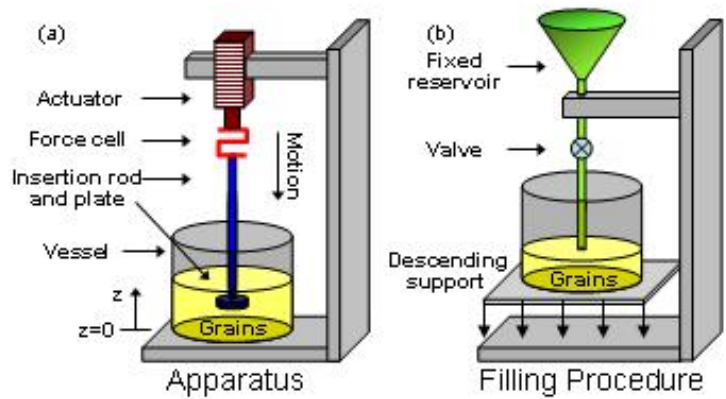

Figure 1, M. B. Stone *et al.*

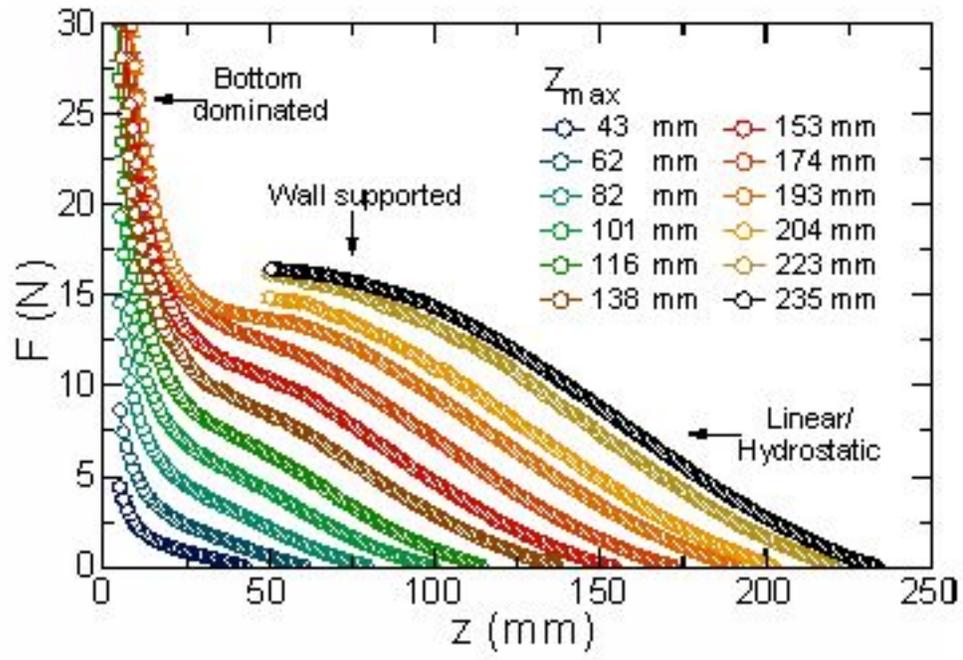

Figure 2, M. B. Stone *et al.*

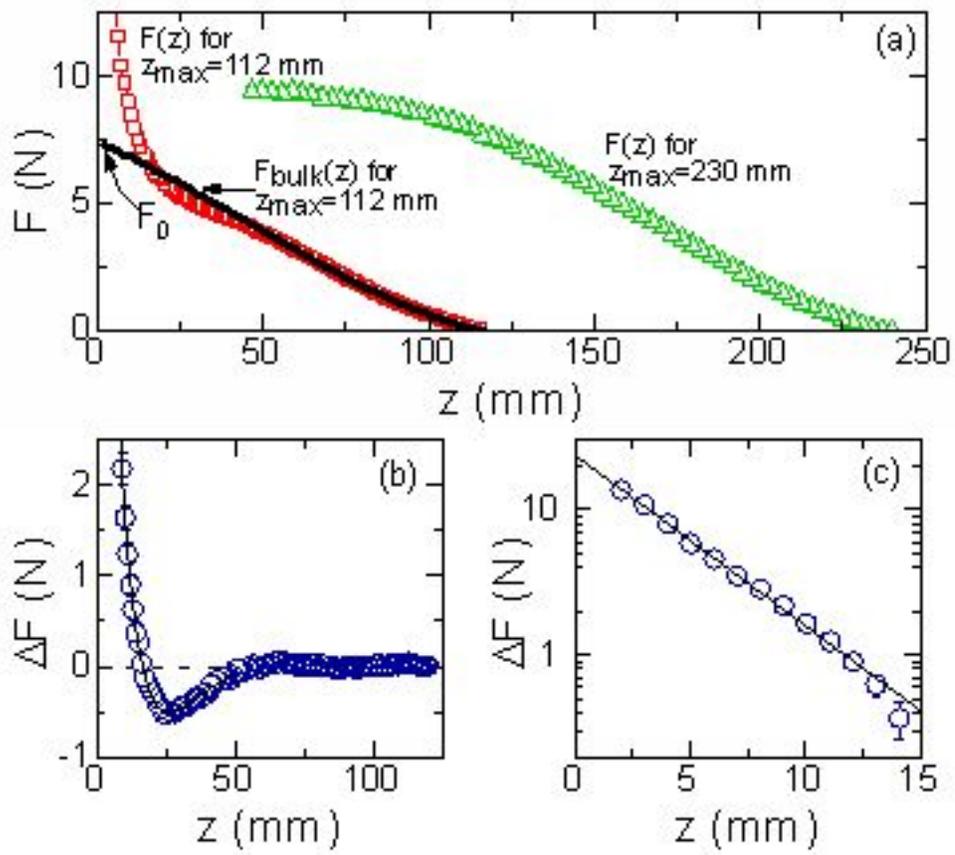

Figure 3, M. B. Stone *et al.*

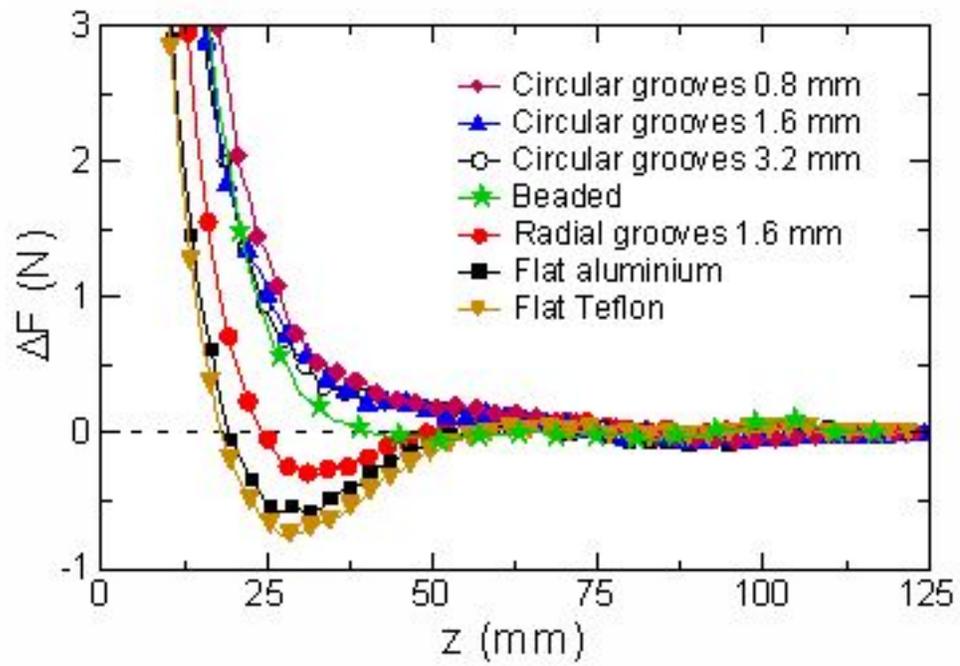

Figure 4, M. B. Stone *et al.*

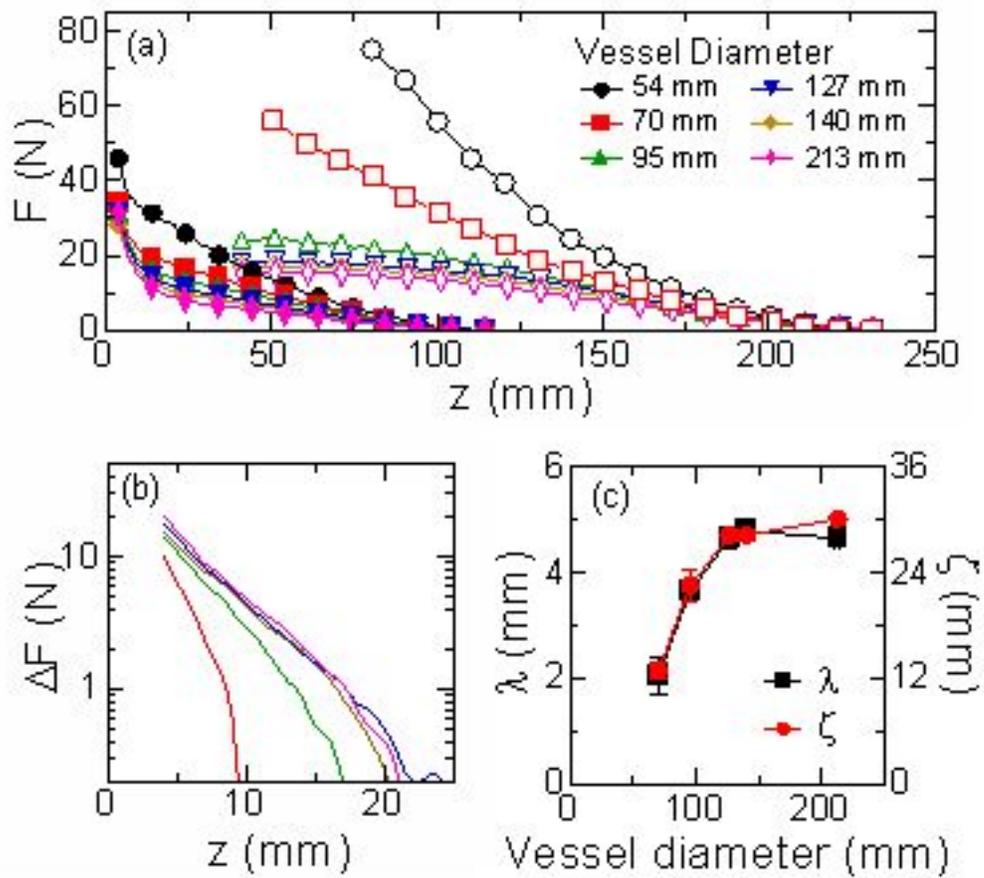

Figure 5, M. B. Stone *et al.*

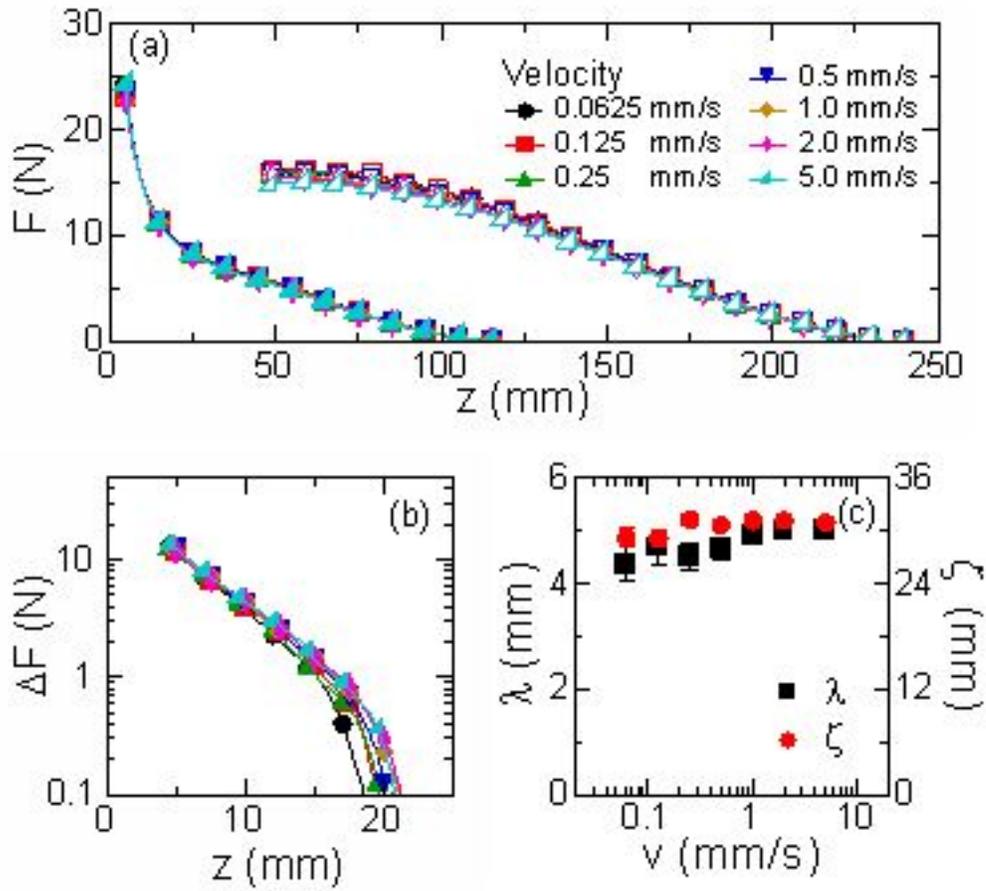

Figure 6, M. B. Stone *et al.*

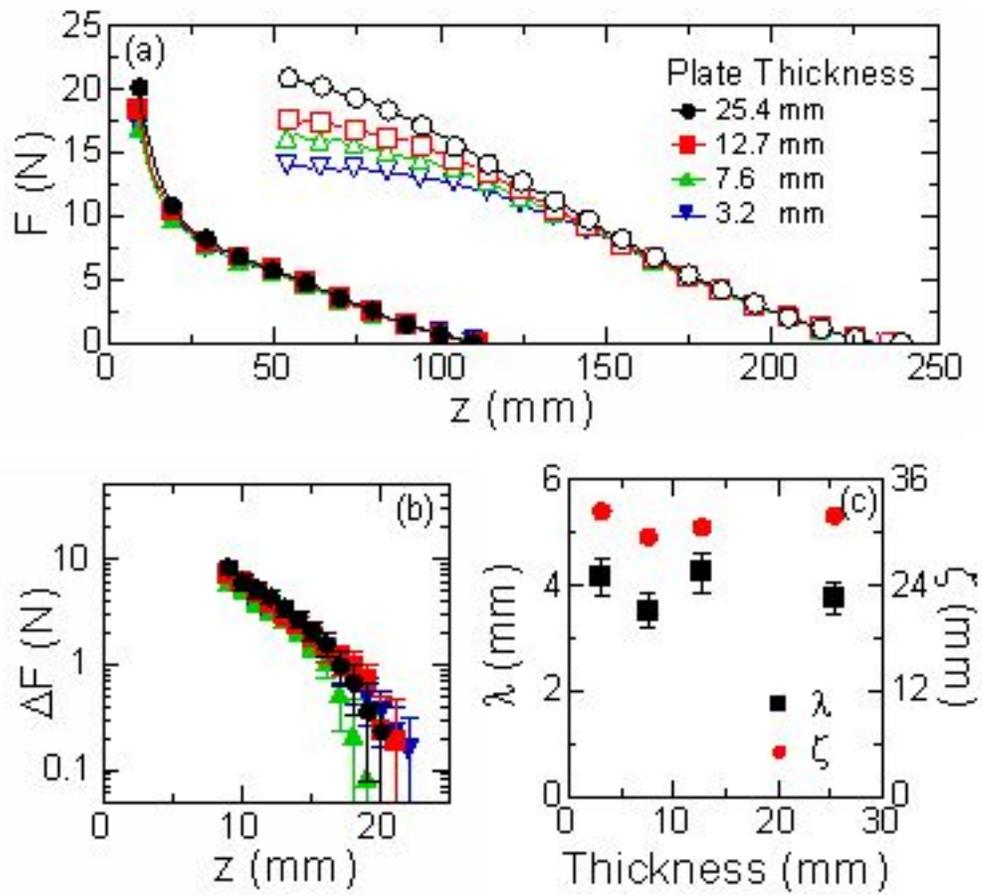

Figure 7, M. B. Stone *et al.*

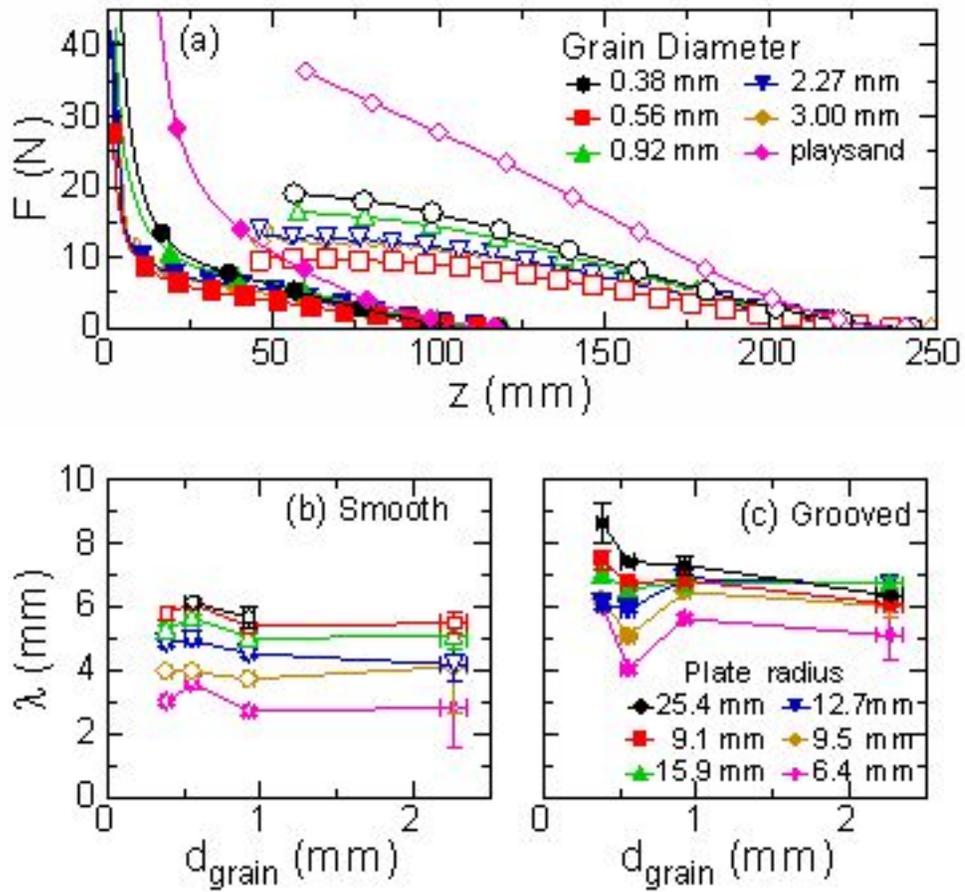

Figure 8, M. B. Stone *et al.*

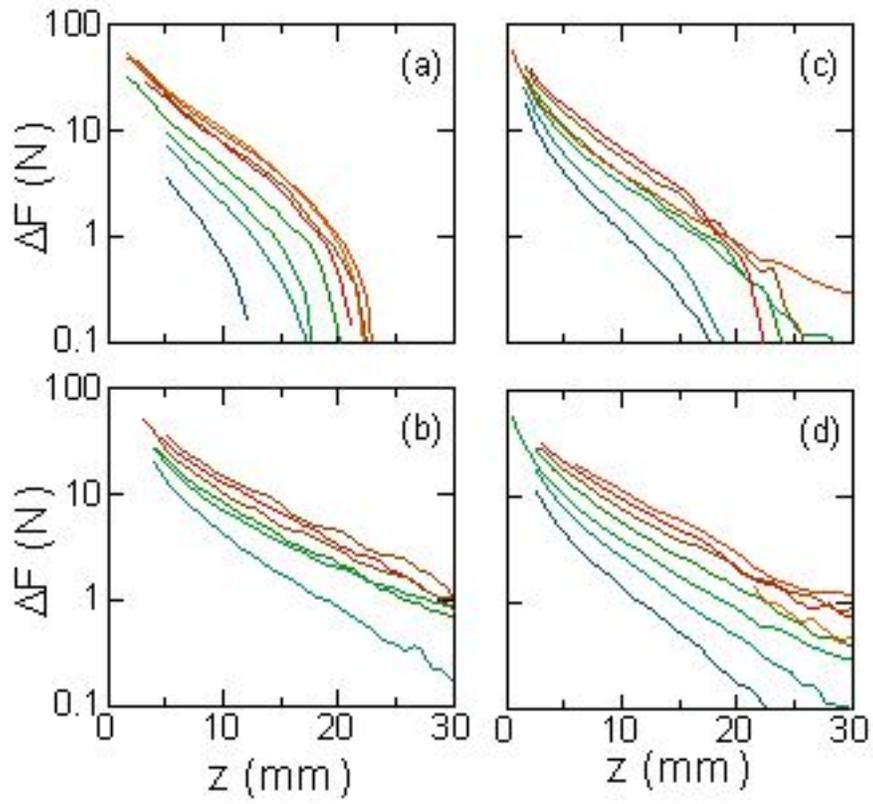

Figure 9, M. B. Stone *et al.*

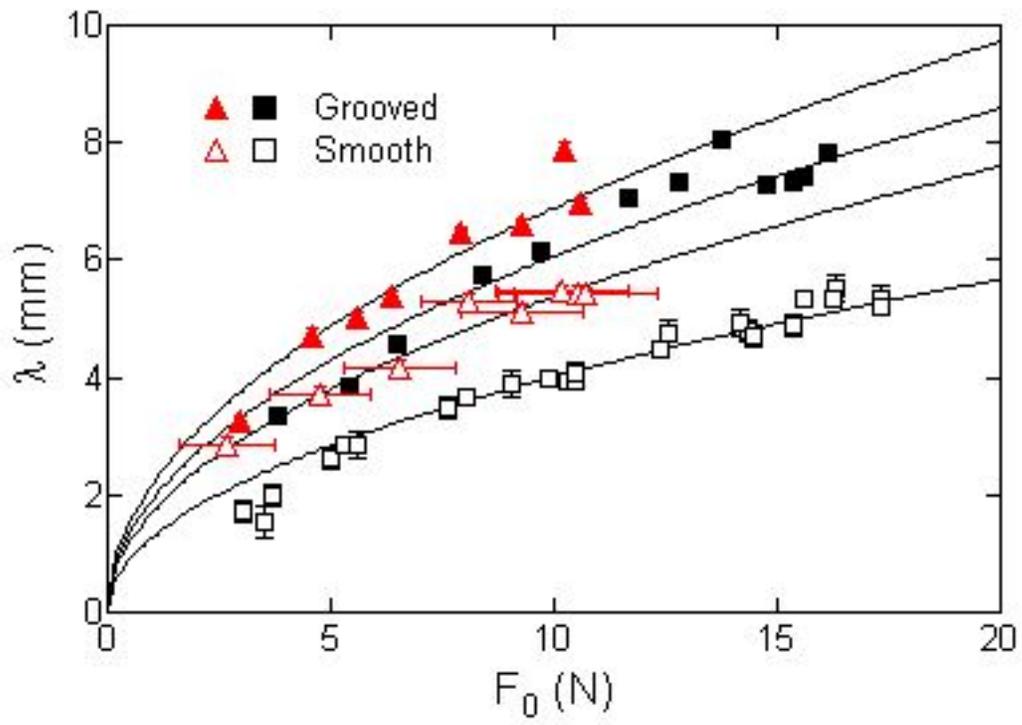

Figure 10, M. B. Stone *et al.*

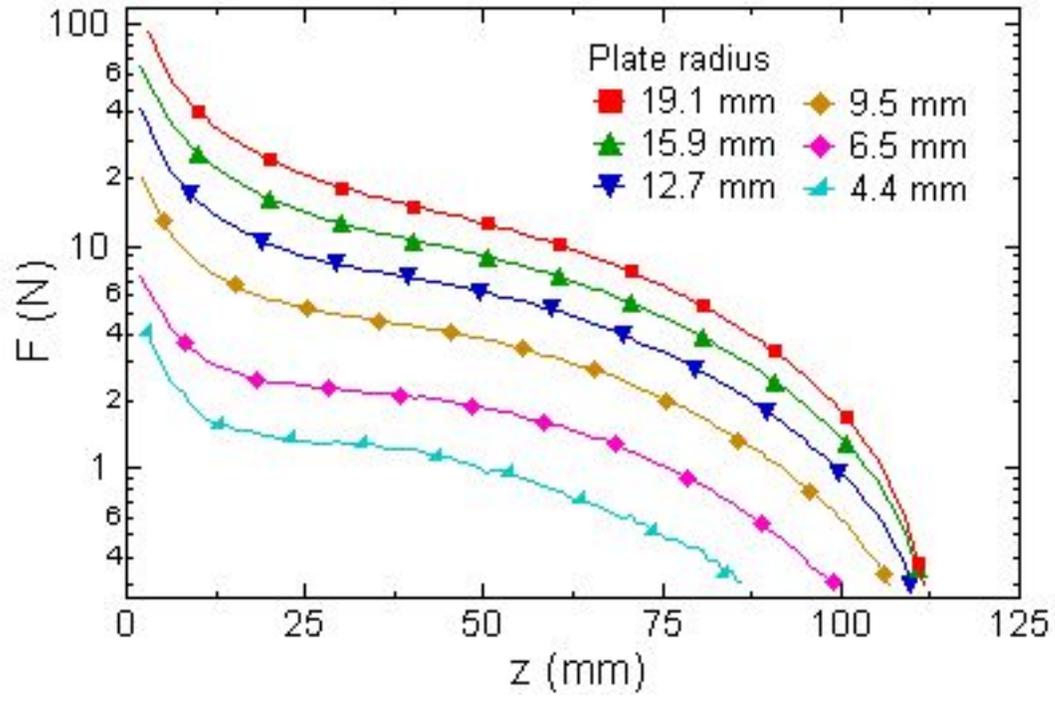

Figure 11, M. B. Stone *et al.*

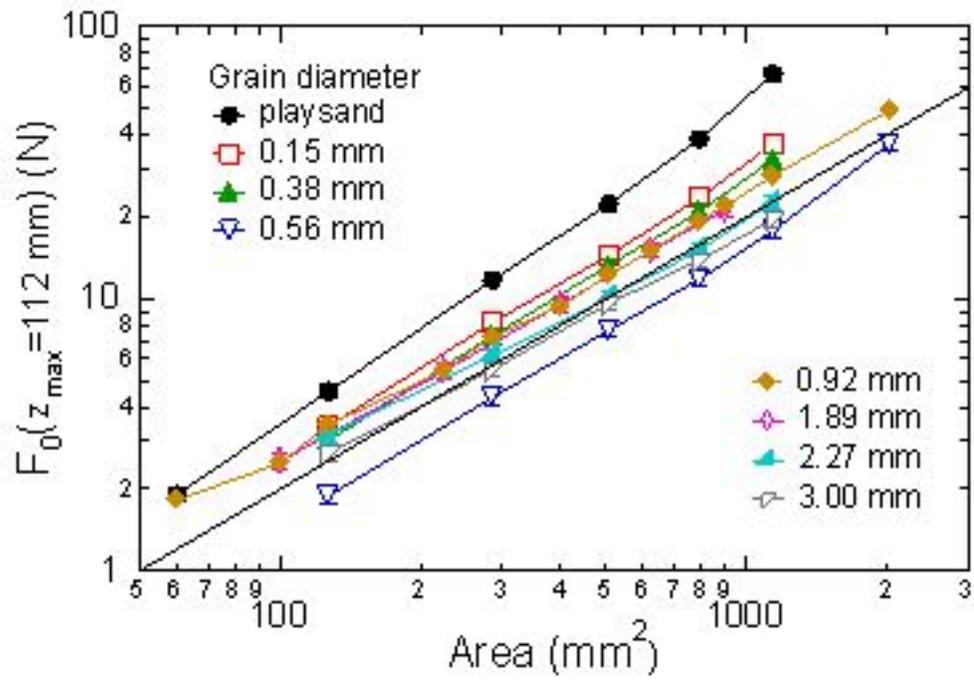

Figure 12, M. B. Stone *et al.*

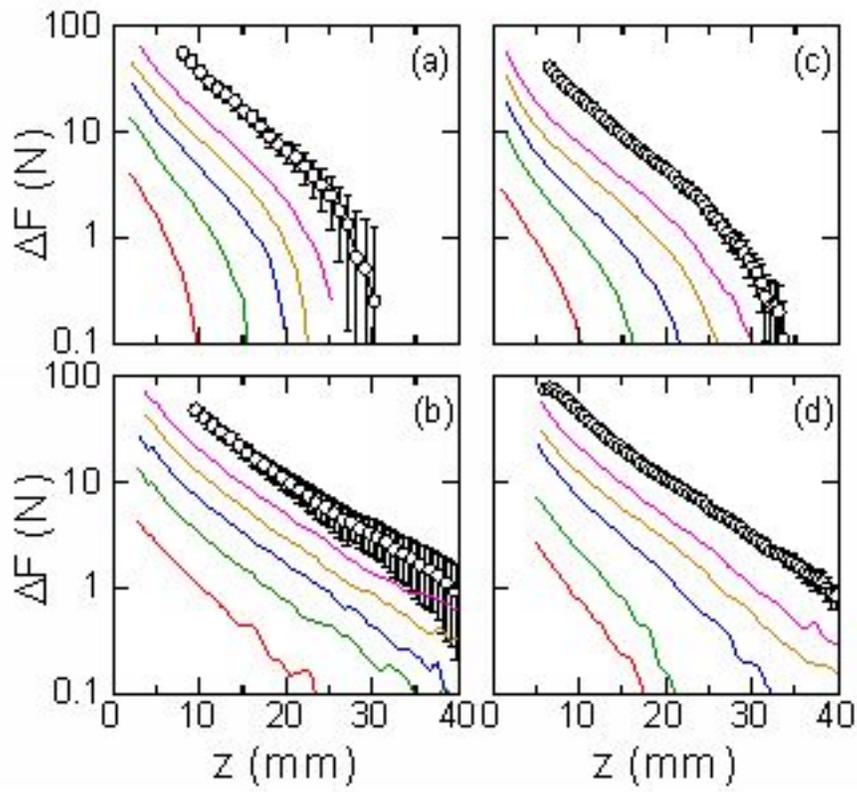

Figure 13, M. B. Stone *et al.*

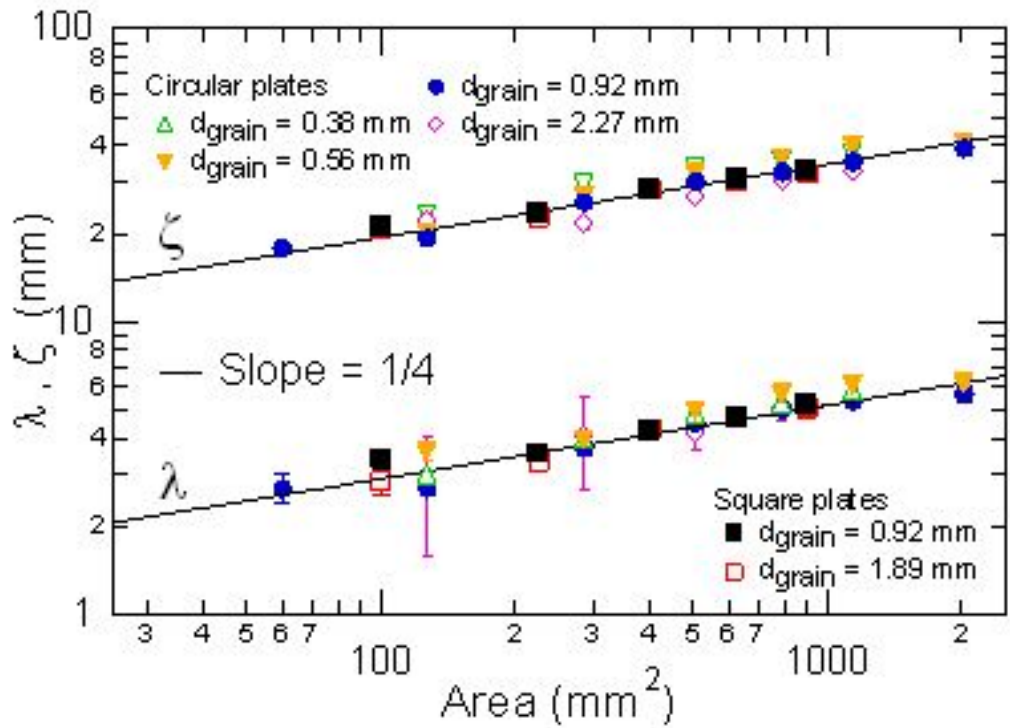

Figure 14, M. B. Stone *et al*

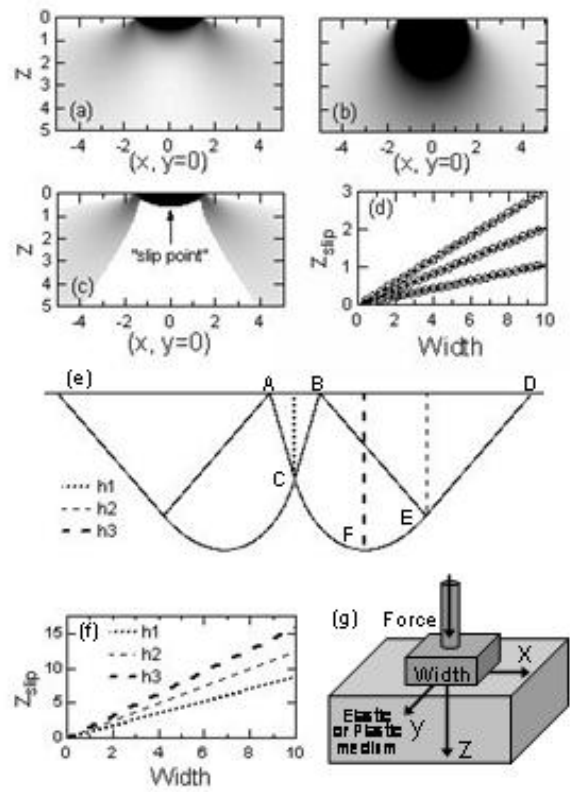

Figure 15, M. B. Stone *et al.*

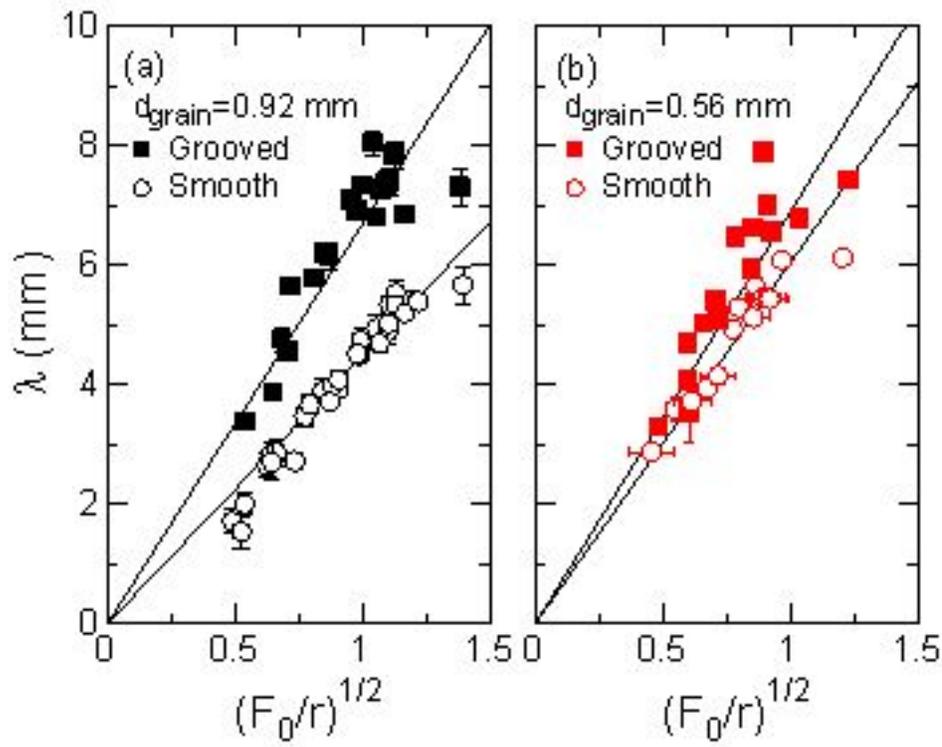

Figure 16, M. B. Stone *et al.*

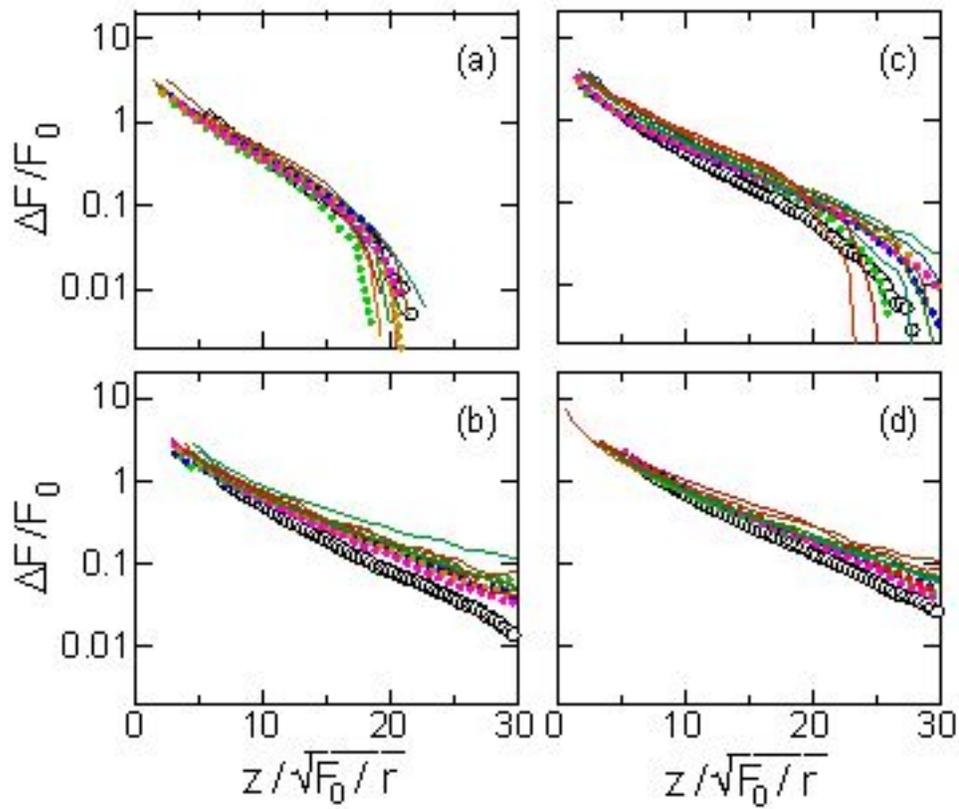

Figure 17, M. B. Stone *et al.*